\documentclass[manuscript,screen]{acmart}

\AtBeginDocument{%
  \providecommand\BibTeX{{%
    \normalfont B\kern-0.5em{\scshape i\kern-0.25em b}\kern-0.8em\TeX}}}

\setcopyright{none}

\begin{document}

\title[The craft and coordination of data curation]{The craft and coordination of data curation: complicating "workflow" views of data science}



\author{Andrea K. Thomer}
\email{athomer@umich.edu}
\orcid{0000-0001-6238-3498}
\affiliation{%
  \institution{School of Information, University of Michigan}
  \city{Ann Arbor}
  \state{Michigan}
  \country{USA}
}
\author{Dharma Akmon}
\affiliation{%
  \institution{ICPSR, University of Michigan}
  \city{Ann Arbor}
  \state{Michigan}
  \country{USA}
}
\author{Jeremy York}
\affiliation{%
  \institution{School of Information, University of Michigan}
  \city{Ann Arbor}
  \state{Michigan}
  \country{USA}
}
\author{Allison R. B. Tyler}
\affiliation{%
  \institution{School of Information, University of Michigan}
  \city{Ann Arbor}
  \state{Michigan}
  \country{USA}
}
\author{Faye Polasek}
\affiliation{%
  \institution{School of Information, University of Michigan}
  \city{Ann Arbor}
  \state{Michigan}
  \country{USA}
}
\author{Sara Lafia}
\affiliation{%
  \institution{ICPSR, University of Michigan}
  \city{Ann Arbor}
  \state{Michigan}
  \country{USA}
}
\author{Libby Hemphill}
\orcid{0000-0002-3793-7281}
\affiliation{%
  \institution{School of Information \& ICPSR, University of Michigan}
  \city{Ann Arbor}
  \state{Michigan}
  \country{USA}
}
\author{Elizabeth Yakel}
\affiliation{%
  \institution{School of Information, University of Michigan}
  \city{Ann Arbor}
  \state{Michigan}
  \country{USA}
}

\renewcommand{\shortauthors}{Thomer, et al.}

\begin{abstract}
Data curation is the process of making a dataset fit-for-use and archiveable. It is critical to data-intensive science because it makes complex data pipelines possible, makes studies reproducible, and makes data (re)usable.
Yet the complexities of the hands-on, technical
and intellectual work of data curation is frequently overlooked or downplayed. Obscuring the work of data
curation not only renders the labor and contributions of the data
curators invisible; it also makes it harder to tease out the impact
curators' work has on the later usability, reliability, and
reproducibility of data. To better understand the specific work of data
curation -- and thereby, explore ways of showing curators' impact -- we conducted a close examination of data curation at a large
social science data repository, the Inter-university Consortium of Political and Social Research (ICPSR). We asked, What does curatorial work entail at ICPSR, and what work is more or less visible to different stakeholders and in different contexts? And, how is that curatorial work coordinated across the organization? We triangulate accounts of data curation from interviews and
records of curation in Jira tickets to develop a rich and detailed
account of curatorial work. We find that curators describe a number of craft practices needed to perform their work, which defies the rote
sequence of events implied by many lifecycle or workflow models. \textbf{Further, we show how best practices and craft practices are deeply intertwined.}
\end{abstract}

\ccsdesc[300]{Human-centered computing~Collaborative and social computing systems and tools}
\ccsdesc[300]{Applied computing~Document preparation}
\ccsdesc[300]{Information systems~Digital libraries and archives}

\keywords{data curation, knowledge infrastructure, craft, coordination, workflows, social science data}

\maketitle

\section{Introduction}
\label{introduction}
Data curation -- the technical work put into datasets to make them fit-for-use and accessible over the long-term -- is critical to data-intensive science \cite{Borgman2019-re,Faniel2011-lb,Muller2019-gi,Palmer_Carole_L_Weber_Nicholas_M_Munoz_Trevor_Renear_Allen_H2013-ll,Hey_T_Tansley_S_Tolle_K2009-hv}. In data science contexts, this work is often referred to as munging, wrangling, or processing, with a particular focus on the working data into a usable format; this work of making data fit-for-use can take up to  80\% of a data scientists’ day-to-day work \cite{Wickham2014-aa}. In institutional contexts -- for instance, in large scientific data archives or institutional repositories -- data curation is likely to involve the application of standards in addition to data munging, with a particular focus in making data shareable and easy to reuse. In both contexts: the ways in which data are transformed and manipulated prior to analysis have significant impacts on the quality and reliability of a study \cite{DIgnazio2020-es, National_Academies_of_Sciences_Engineering_and_Medicine2019-dh,Borgman2015-ap}. Additionally, making data ready to archive or share is increasingly required by both funding agencies and journals.

Yet, despite its importance, data curation -- and data curators -- are often overlooked in accounts of data science. Job ads for data scientists frequently call for data ``unicorns,'' ``ninjas,'' and ``rock stars'' to wrangle messy datasets through mythic abilities (not through skill or craft), or ``janitors'' as if data processing were a rote sanitizing process that requires little specialized expertise or training \cite{DIgnazio2020-es}. Data science clients similarly think of data work as ``magic'' \cite{Kross2021-gz} -- which, while seemingly complimentary, obscures the skill and effort needed for this work. In addition to obscuring the skill needed for curatorial work, this invisibilization obscures the varied judgements,  decisions, and data processing steps that go into data processing and have big impacts on the final trustworthiness, reproducibility and auditability of a study.  

The work of data curation can also be obscured, somewhat ironically, through attempts to render it visible as part of a regularized workflow. Workflows and curatorial best practices aim to break curation into a discrete set of steps, or show it as one 'phase' of work in a project (e.g. \citet{Muller2019-gi,Higgins2008-og}). The goal of workflow representations is to make curation more reproducible and routine -- but it comes at the cost of obscuring the skill needed to do these tasks well and furthering the idea that curatorial work is a rote task that just any human can be plugged into. 

Obscuring data curation also renders the labor and contributions of data curators invisible \cite{Plantin2019-fs}. Like other forms of service work, well-executed curation is hidden \cite{Suchman1995-zr}. Additionally, obfuscating curatorial work makes it challenging to understand the impact of specific curatorial actions, and therefore to efficiently prioritize, plan, or fund data curation (anonymized for review). Without understanding the impact of data curation, the developers of curatorial tools cannot assess or prioritize which features and functionalities will best increase curatorial efficacy or later data reuse.   
 
To better make the work of data curation visible, we conducted a close examination of data curation at a large social science data archive, the Inter-university Consortium for Social and Political Research (ICPSR). ICPSR recently adapted external standards and best professional practices to create robust internal guidelines for curation, and the scale, centrality, and collaborative aspect of curatorial work at ICPSR make it an excellent site for a case study of data curation. ICPSR is the largest social science data archive in the world, and it contains datasets from over 16,000 studies. ICPSR's professional, in-house curation activities distinguish it from other data repositories such as the UCI Machine Learning Repository\footnote{https://archive.ics.uci.edu/ml/index.php} or Dataverse\footnote{https://dataverse.harvard.edu/} where data providers are expected to curate data themselves. 

This research is part of a larger project focused on understanding the impact of curatorial work, and aimed at developing metrics that better measure and account for the benefits of that work. We aim to make curatorial work more visible, and thereby easier to account for in budgets and in academic promotion cases. Our methods include interviews with ICPSR stakeholders, as well as computational analysis of curation logs. 
Here, 
we address the following research questions:

\begin{enumerate}
    \item What does curatorial work entail at ICPSR, and what work is more or less visible to different stakeholders and in different contexts?
    \item How is that curatorial work coordinated across the organization?
\end{enumerate}
We drafted these questions with the goal of understanding both the visible and invisible work that goes into data curation; prior studies have shown that much of this work escapes view \citet{Plantin2019-fs}, but fewer have sought to specify what, exactly, is invisible.

By triangulating accounts of data curation from interviews and records
of curation in Jira tickets, we develop a rich and detailed account of
curatorial work at ICPSR. In doing so, we bridge research in CSCW and the library and archival sciences on data curation. We find
that while there are several standard curatorial activities performed at ICPSR, and well defined standards for different "levels" of curation, considerable craft and coordination are needed to do this work well; in other words, craft is needed to "work" a workflow. This non-technical work is necessary to facilitate technical work, and has been less
well-defined in prior discussions of data curation. Surfacing the role of craft and coordination has important
implications for curatorial projects' and teams' planning. This defies the rote sequencing of events implied by
many lifecycle or workflow models. We provide a detailed account of
curatorial workflows at ICPSR and explain how workflow-based accounts of
data curation can obscure both the individual  skilled ``artistry'' and coordination 
necessary in this work.

We additionally reflect on the visibility of data curation, both within ICPSR and to data users. As \citet{Plantin2019-fs} has previously
described, much of data curators' work is intentionally kept invisible
to the final data consumers -- yet, curators experience their jobs as
being hypervisible to their supervisors, via the extensive documentation
they create. We discuss how
different kinds of invisible work are at play in data curation at ICPSR
and explain how CSCW and data science can benefit from better
understanding the judgements and skill that goes into effective data curation.

\section{Prior work}
\label{prior-work}

\subsection{What is data curation?}

For the purposes of our research at ICPSR, we have defined data curation as the work needed to make a dataset fit-for-use over the long term (anonymized for review) 
A detailed description of data curation activities at ICPSR is provided in Sections~\ref{research-site} and~\ref{triangulating-with-jira-tickets} . Depending on the scholarly community, researchers have described this kind of work in varying ways and employed several research data management-related terms as synonyms. 

Much of the CSCW literature discusses curation in terms of fitness for use aimed at a single user or end goal \cite{Taylor2015-bi,Kandel2011-ss,feger2020}. For example, Feger et al. focus on the activities “essential for generating reproducible artefacts,” while Kandel et al. are concerned with “wrangling” data to enable meaningful analysis for the research at hand. Others, such as \cite{Muller2019-gi} define curation as a type of human “intervention” that includes data-cleaning, converting metadata, and data alignment. In proposing ``Datasheets for datasets'', Gebru and colleagues [\citeyear{Gebru2019-td}] argued that data used in machine learning should carry documentation that describes its collection processes and transformations, two steps in the process of curating data for reuse. For data science workers, data curation is a collaborative activity centered on information exchange and data and code transparency \cite{Zhang2020-bf}. In their ethnographic study of the Long Term Ecological Research Network, \citet{Karasti2006-lh} describe “information managers’” data stewardship strategies and strengths, including their ability to turn localized, heterogeneous data into a networked resource. In this way, data curation, and the development of information infrastructures, is a long-term sociotechnical endeavor.

Libraries, archives, and data repositories conceptualize the aims of curation more broadly, emphasizing supporting data’s long-term preservation and usability for a multitude of potential future purposes. Researchers in this area have defined digital curation as “the active involvement of information professionals in the management, including the preservation, of digital data for future use” \cite{Yakel2007-lp}, while \emph{data curation} refers to active management throughout the data lifecycle \cite{Palmer_Carole_L_Weber_Nicholas_M_Munoz_Trevor_Renear_Allen_H2013-ll}.\footnote{“Digital curation” is often used as a broader, more general term for management of any collection of digital objects, whereas “data curation” refers to the long-term care and management of data specifically. We focus on data curation in this paper, but draw on research on digital curation and preservation where relevant.} Models of data curation, such as the Digital Curation Centre’s Curation Lifecycle Model \cite{Higgins2008-og} and the Big Data Lifecycle Model \cite{Pouchard2016-gz}, identify different curation needs at different phases of the lifecycle of data and provide a workflow for curation work performed initially by data producers and then by the data curators and processors at archives and repositories. Other models take the form of terminological frameworks, such as the “Data Practices Vocabulary” by  \citet{Chao2015-sk}, which outlines a taxonomy of terms describing curatorial work. There are overlaps between CSCW and LIS/archival conceptualizations of data curation; the differences tend to be in the detail in which curatorial work is described and the focus on a long-term future for the data in LIS/archival science. For instance, \citet{Higgins2008-og} description of the entire curation lifecycle mirrors the five stages of data science work practices by \citet{Muller2019-gi}. Higgins' model includes 8 sequential phases of work with data: Conceptualize, Create/Receive, Appraise/Select, Ingest, Preservation Action, Store, Access Use and Reuse, and Transform. Muller et al.’s data science lifecycle model includes 5 sequential phases of work, with curation in the center: Discovery, Capture, Curation, Design, and Creation (2019). 

\subsection{What renders data curation invisible?}
What may not be clear from all these definitions of curation work is that when curation is done well, it is invisible to the data’s users. Successful data curation enables reusers to readily access and use datasets and does not highlight the data transformations or metadata generation that make the data ready for use. Curation work can be done by the data producers themselves as part of their research data management process. This curation work includes cleaning, organizing, and storing their data for their own localized research needs \cite{Wallis2008-cs}. Often, however, curation tasks—the data manipulation, cleaning, documentation, preservation, and other work discussed below—are carried out by data curators or processors within the repository or archive that has selected the data for inclusion in its holdings \cite{Johnston2018-gy}. 

The role of the data archive is two-fold: to enable the researcher to share their data with the scientific community, and to ensure the data’s long-term accesibility and preservation \cite{Green2007-bj}. The data producer deposits their data with the archive, and then at some future moment, the data appear in a standardized form, ready for use, with inconsistencies smoothed over and issues addressed, presented to the data user without the explicit traces of the curation work that are only visible to those within the data archive \cite{Plantin2019-fs,Kervin2014-ld}. By producing data according to professional standards, for instance, curators purposefully render themselves and their work invisible \cite{Plantin2019-fs}.

The invisibility of the work makes it hard to classify. Prior scholars have explained myriad ways that work can be invisible. For instance, Nardi and Engeström [\citeyear{Nardi1999-tu}] identify four types of invisibility at work: 1) work done in invisible places, such as the highly skilled behind-the-scenes work of reference librarians; 2) work defined as routine or manual that actually requires considerable problem solving and knowledge, such as the work of telephone operators; 3) work done by invisible people such as domestics; and 4) informal work processes that are not part of anybody’s job description but which are crucial for the collective functioning of the workplace, such as regular but open-ended meetings without a specific agenda, informal conversations, gossip, humor, storytelling.

Similarly, \citet{Star1999-fx} propose three forms of invisible work: where “the act of working or the product of work is visible to both employer and employee, but the employee is invisible”; where the “workers themselves are quite visible, yet the work they perform is invisible or relegated to a background of expectation”; and, when “both work and people may come to be defined as invisible” according to particular indicators. Curators possess different types of invisibility. For example, \citet{DIgnazio2020-es} recognize the highly skilled behind-the-scenes work prevalent in what they term ``data cleaning'' at the same time recognizing that others discount the intellectual work required. \citet{Kross2021-gz} report on the black-boxing of curation that leads clients to deem the results ``magic.'' 

In social computing, curation work is sometimes invisible because it occurs during data collection or generation. Machine learning, computer vision, and social media studies often use ``found'' data \cite{Hemphill2021-ot, Paullada2021-ah, Jo2020-mu} and render curatorial decisions such as ``what data should be available,'' ``in which format(s) should data be provided,'' or ``how should this data be sampled'' invisible. For instance, datasets scraped from the web (such as Flickr photos \cite{Zhang2015-eh,Scheuerman2021-xo} or Wikipedia talk pages \cite{Wulczyn2016-ws,Wulczyn2017-at}) suffer from biases in representation \cite{Jo2020-mu}. The kinds of curation activities that occur in archives could address those biases by adjusting samples, weights, or documentation. Annotation processes in which humans add labels to data that can then be used in machine learning tasks (e.g., facial recognition, hate speech detection), are another data generation step that are often minimized in reports about the research that depend on them. For instance, \citet{Scheuerman2021-xo} explain that reference datasets used in computer vision tasks in papers   are not well-described, and the details of the annotation process and potential biases introduced are missing. They argue that the value of ``efficiency'' is responsible for this pattern and that explicitly working toward other values such as ``care'' could improve data curation practices in computer vision.

\subsection{Craft in data work}

Throughout the CSCW literature, there has been discussion of the craft needed in technical work; recent papers have begun to apply this framework more specifically to work with data. \citet{Barley1997-cc}'s well known volume on the topic argues that ``technical work sits at the intersection of craft and science, combining attributes of each that are normally thought to be incompatible." These attributes include the use of complex technologies; a reliance on contextual knowledge and skill; the development of abstract conceptual representations to guide work; and a grounding in a community of practice \cite{Barley2020-cf}. \citet{Rosner2018-uq} argue that appreciation of craft work in computer science, though, is marred by “gendered narratives” about the value of such labor, which, “both haunt and inform HCI’s ideas of technological belonging, participation, and differentiation." 
 In the context of data work, scholars have focused on how craft practices are used to process and interpret data. \citet{Mentis2016-iw} examine surgeons collaborating remotely over image data to craft a shared interpretation. More recently, \citet{Muller2019-gi} view the data science pipeline process as one of crafting the data, in which workers combine technical skill, expertise working with abstraction and representations and decision-making to accommodate unexpected issues and application of more routine techniques and automated scripts.
 
Within the information sciences, discussion of craft has largely focused on its role as part of librarianship and archival practice. Archivist Trevor Owens brings these conversations forward to a digital context, in his discussion of digital preservation (an aspect of data curation) as a craft, “best understood as part of an ongoing professional dialog on related but competing notions of preservation that goes back to the very beginnings of our civilizations” \citet{Owens2018-gm}. He further writes, 
\begin{quote}
``digital preservation must be a craft and not a science because its
praxis is; 1) grounded in an ongoing and unresolved dialog with the
preservation professions and 2) it must be responsive to the inherent
messiness and historically contingent nature of the logics of
computing."
\end{quote}

When this craftful work is done collaboratively, considerable articulation work and coordination are needed to do it successfully. Articulation work, “consists of all the tasks needed to coordinate a particular task, including scheduling subtasks, recovering from errors, and assembling resources” \cite{Gerson1986-ng}, whereas coordination is the “process expertise” entailed in said scheduling and assembly \cite{Barley2020-cf}. Articulation and coordination work have been shown to be critical in data curation for multiple reasons. They are needed in maintaining a knowledge infrastructure’s stability \cite{Karasti2004-vq}; facilitating the selection and enactment of data curation protocols \cite{Darch2020-dd}; refactoring data structures and vocabularies \cite{Thomer2018-oq}; supporting infrastructure design \cite{Baker2007-yg}; enabling navigation of information during the process of scientific discovery \cite{Palmer2007-ok,Palmer2007-ef}; and are a core component of the “data labours” of building and sustaining data collections \cite{Nadim2016-rm}. \citet{Erickson2016-ox} describe the articulation work needed by knowledge workers, such as data curators, to configure infrastructural solutions to overcome technical and contextual constraints in tools and workplaces. A recurrent theme in these papers is the lack of tools to support this coordination and articulation work; curators must coordinate their work often in spite of these tools, rather than through them. 

\section{Methods}
\label{methods}
In this paper, we report on a mixed methods study to examine different aspects of the
data curation process. We leverage two bodies of data: 1)
semi-structured interviews with stakeholders across ICPSR; and 2) records of curation work in Jira
tickets, a subset of the internal ICPSR documentation
that records data curators' work.

\subsection{Research Site}
\label{research-site}
ICPSR, founded in 1962, is one of the oldest and largest curated social science data
archives in the world. It not only collects, curates, and disseminates data in a
broad range of disciplines including political science, sociology,
demography, education, criminology, public health, among others, but it
is also a leader in repository infrastructure, data curation standard
setting, and innovation in data curation. ICPSR's archives include over
16,000 studies containing nearly 6 million variables. ICPSR's
collections are organized into separate archives representing different
subject areas and often sponsored by federal agencies and foundations.
We selected ICPSR for three main reasons: 1) curation processes are well
articulated and documented; 2) the volume of data curation is large
enough for patterns to emerge, and 3) we were given access to both
documentation and staff to conduct an in-depth study of the curation
process.

ICPSR's organizational structure also makes it possible to study data curation in depth.
Several years ago, ICPSR centralized curation into one unit. Curators previously worked for individual archives within ICPSR, reporting to a project manager, who, in turn, reported to an archive director; now curation staff, project management staff, and archive directors sit within their own distinct organizational units (e.g. the Curation unit, the Project Management unit). Part of this re-organization also involved a redesign of curatorial standards. As of 2018, datasets are assigned to one of three standard "level"s of curation which articulate specific curatorial actions that vary according to the amount, intensiveness, and complexity of effort required as well as the end product delivered. These levels provide a standard for curation actions and expected outputs, which are assigned based on the format, size, and level of preparation performed by the data creator prior to deposit \cite{icpsr-curation}. Higher levels of curation are intended to improve the usability of data products. All data deposited with ICPSR receive a base level of curation (``Level 1 Curation"), meaning that curators remediate personally identifiable (disclosive) information and create a metadata record, a Digital Object Identifier (DOI), statistical files, a webpage, and a codebook explaining the variables in the data collection. ``Level 2 Curation" includes all ``Level 1" actions, plus additional data transformations, completeness checks, and preparation of the data for online analysis. ``Level 3 Curation" is intensive and includes custom documentation, attaching survey question text to variables, and indexing variables for search. Non-tablular data, such as qualitative or spatial data, typically require ``Level 3" curation. For example, the ``TransPop, United States, 2016-2018" study shown in Figure~\ref{fig:jira-ticket} is curated at Level 3, meaning that additional curatorial tasks have been assigned and more time has been budgeted for intensive curation including extensive disclosure review and remediation and creating searchable question text.

\subsubsection{Interviewees and semi-structured
interviews}
\label{interviewees-and-semi-structured-interviews}
The internal stakeholders in ICPSR curation extend beyond the
curation unit itself. In order to better understand the impact of
curation within the data repository, we conducted in-depth,
semi-structured interviews with 37 ICPSR stakeholders comprising six
staff groups: archive directors, project managers, curation supervisors,
curators, user support, and bibliographers. Each archive is led by a
director who spearheads collection development efforts, secures funding,
interacts with archive sponsors, and attends disciplinary conferences and meetings to
expand the reach of the archive. When ICPSR ingests data, a project
manager shepherds the data through curation and dissemination, serving as a conduit between the curators and the data producers. The
project manager works with the archive director and the curation
supervisor to determine which curation activities to apply to the data
and how to prioritize the data relative to other studies in the queue. User support personnel bridge between data reusers and either project managers or curators as questions about the data arise. Bibliographers track use of the all ICPSR's curated datasets and maintain an extensive bibliography of that use.

Curation work is accomplished primarily by a dedicated team of curators
(n = 32) and their curation supervisors (n = 5). We note that data curators are typically entry level employees; this is not always the case at data archives. ICPSR requires curators
to have experience with statistical software (e.g., SPSS, Stata), data
preparation, and social science research methods. ICPSR actively curates
data to ensure that they comply with the FAIR principles (i.e., are findable,
accessible, interoperable, and reusable)
\cite{Wilkinson2016-rg}. Generally, curators review data
for sensitivity and re-identification risk, generate metadata
\cite{Vardigan2008-gc}, identify missing values, index
variables for future search and discovery, link question text to
variables, apply subject terms to the study, and generate multiple
formats (e.g., SPSS, Stata, plain text) of the data files. Curators also
pass along citations to publications that use the data to the
bibliographers for inclusion in the ICPSR Bibliography of Data-Related
Literature. These tasks are completed by individual curators and
reviewed by their supervisor or a senior curator before disseminating
the data for reuse. On an ongoing basis, the bibliographers also search
for additional citations to studies archived at ICPSR.

The interviewees were selected using purposive sampling
\cite{Miles2014-jh}; we requested interviews with all
personnel working in the specific roles identified. The only
criteria used to filter out potential respondents was for the curators
themselves: due to the time required to become familiar with curatorial
work, we limited our interview requests to those curators who had been
working for at least one year so that they had built up some
expertise in curation work and could better reflect on the processes.
The interviews focused on understanding how the different
stakeholder groups measured the value and impact of curation work (see Appendix \ref{appendix-a-interview-protocol} for our full interview protocol).

We conducted 37 semi-structured interviews with archive staff that
enabled us to probe further into the responses and ask questions that
were specific to each role
\cite{Dearnley2005-zf,Hesse-Biber2005-zp,Rubin2012-tn}.
Interviews were conducted in 2019 and 2020; 12 were
face-to-face interviews conducted before the COVID-19 pandemic led our institution (anonymized for review) to transition to remote work, and the
remaining 25 were conducted remotely using the Zoom and Google Meet
platforms. Table \ref{tab:stakeholders} details our interview participants. The interviews
were recorded, and then transcribed by the REV transcription service
and verified. We anonymized our interview transcripts by assigning all
participants identifiers. Our study was reviewed by our university's Institutional Review Board and found to be exempt from on-going
oversight IRB information anonymized for review].

We began analysis using a deductive approach, using a "start list" of codes derived from our research questions, interview questions, and our knowledge of prior literature. In this first round of codes, we paid particular attention to identifying curatorial actions at ICPSR. Codes were were iteratively expanded and refined
through subsequent rounds of inductive coding \cite{LeCompte2012-dq,Miles2014-jh}. Analysis of the interviews was completed using the qualitative data
analysis program NVivo. Because multiple team members were conducting the coding, we establish inter-rater reliability (IRR) to endure coherence. As we began
establishing IRR between two members of the interview team, we realized
that the interviews between the different stakeholders were divergent
enough that IRR would need to be established within each stakeholder
group. With the exception of the single User Support interview (59.2\%)
and the three Bibliography Team transcripts (69.5\%), IRR was repeated
within each stakeholder group until at least 70\% was achieved using
Scott's pi \cite{Scott1955-iw}. One member of the
interview team coded transcripts across all stakeholder groups, and two
different members established IRR with her on specific sets of
transcripts. 

After each round of coding was completed, team members reviewed coded data, then met as a group to discuss emergent themes. After our first round, we identified the role of craft and coordination as being key to data curation work, and decided to conduct a secondary round of axial coding to deepen our
analyses. Again, IRR was established between team members until at least
70\% was achieved using Scott's pi. The authors reviewed coded data and again met to discuss the codes as a group. Finally, after reviewer feedback, we conducted a third round of coding, this time diving deeper only into codes related to craft in data curation to deepen our analysis. We met again as a group to discuss emergent themes.



\begin{table}
  \caption{Number of interviews by stakeholder category}
  \label{tab:stakeholders}
  \begin{tabular}{lll}
    \toprule
    Stakeholder Category & Number of Interviews & Interview Codes \\
    \midrule
Archive Director & 7 & AD002-AD006, AD008, AD011 \\
Project Manager & 9 & PM025-PM033 \\
Curation Supervisor & 7 & CS001, CS007, CS009, CS010, CS012-CS014 \\
Curator & 10 & CU015-CU024 \\
Bibliography Team & 3 & BT034-BT036  \\
User Support & 1 & US037 \\
  \bottomrule
\end{tabular}
\end{table}

\subsubsection{Triangulating with Jira tickets}
\label{triangulating-with-jira-tickets}

\begin{figure}
    \centering
    \includegraphics[width=\textwidth,height=\textheight,keepaspectratio]{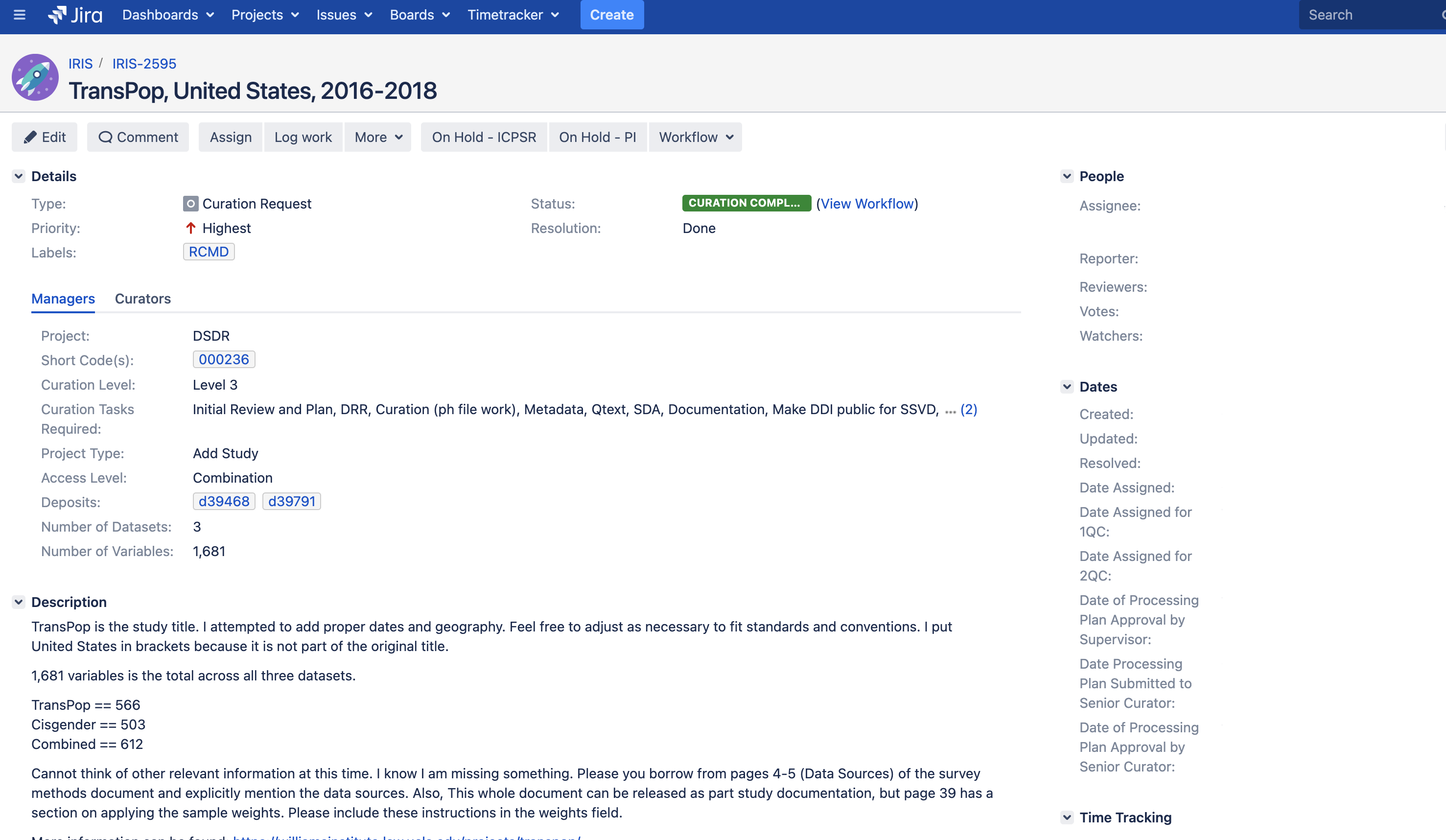}
    \caption{Jira ticket for a single study}
    \label{fig:jira-ticket}
\end{figure}

\begin{figure}
    \centering
    \includegraphics[width=.6\textwidth,height=\textheight,keepaspectratio]{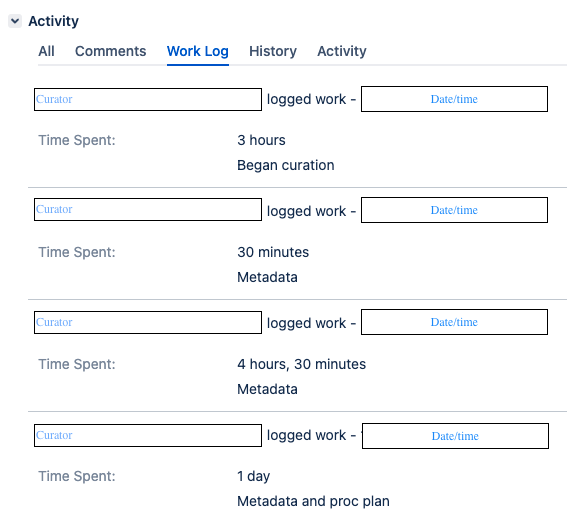}
    \caption{Worklog excerpt from a Jira ticket}
    \label{fig:jira-worklog}
\end{figure}

We triangulated findings from interviews with documentation created
through the curation process, again looking for descriptions of curatorial actions. Curation work at ICPSR is coordinated and
documented across three main sets of documents: processing plans, Jira
tickets, and processing history (PH) files. Jira tickets are the richest and most specific
record of data curators' work. Jira is type of project management software that organizes work through the creation of "tickets" that describe the work that needs doing, and that users can update with progress over time. When data are deposited through the ICPSR deposit system, staff review the data for fit and priority, and a data project manager or assistant generates a Jira ticket (see Figure \ref{fig:jira-ticket}; we removed identifying information from the fields on the right but leave their titles to show what information tickets contain). They provide a study title, the priority of the study, the funder or sponsoring archive, a description of the work curation will need to do, and the level of curation (and any additional tasks) required. Before curation begins work on a Jira ticket, metadata unit staff review the ticket and study metadata. After data project staff and metadata staff approve the ticket, it gets sent to curation for assignment. While curation works on the study, they provide details about their work and progress in the ``worklog'' section of the ticket (see Figure \ref{fig:jira-worklog}). The worklogs offer insights into aggregate time spent on different kinds of curatorial actions at ICPSR.

To classify the parts of worklog descriptions (e.g., "Began curation" and "Metadata and proc plan" in the example in Figure \ref{fig:jira-worklog}), we developed a set of
eight high-level curatorial actions that describe curation work: initial review and planning; data transformation; metadata; documentation; quality checks; communication; non-curation; and other activities (see Figure \ref{fig:taxonomy}). These categories mirrored the codes used in our qualitative analysis. 

We manually coded a randomly selected proportional sample of Jira ticket
worklog entries stratified by curation level. These were coded in
\emph{brat} software \cite{Stenetorp2012-bw} to create labeled
training data to facilitate the automatic classification of the Jira ticket worklogs
(discussed more fully in (anonymized for review)). We trained a computational
model with 0.75 accuracy to assign each worklog entry one of the eight
categories of curatorial actions (summarized in Figure \ref{fig:taxonomy}). For example, a worklog entry ``Discussed curation standards
with supervisor (2 hours)'' is classified as an instance of
``Communication'' while an entry describing ``Recording dataset
limitations in processing notes (10 hours)'' is classified as
``Documentation''. We then aggregated each class of action to analyze
the relative amount of time spent on each.

There are several limitations to this study. First, it documents curation work at one repository. Second, ICPSR is a mature repository with well-articulated policies and procedures. Finally, we did not do direct observation of the curatorial process but rely on direct reports from curation staff and other stakeholders and indirect observation through the Jira tickets.   

\section{Findings: Curatorial work at ICPSR}
\label{findings-curatorial-work-at-icpsr}

\begin{figure}[h]
    \centering
    \includegraphics[width=\textwidth]{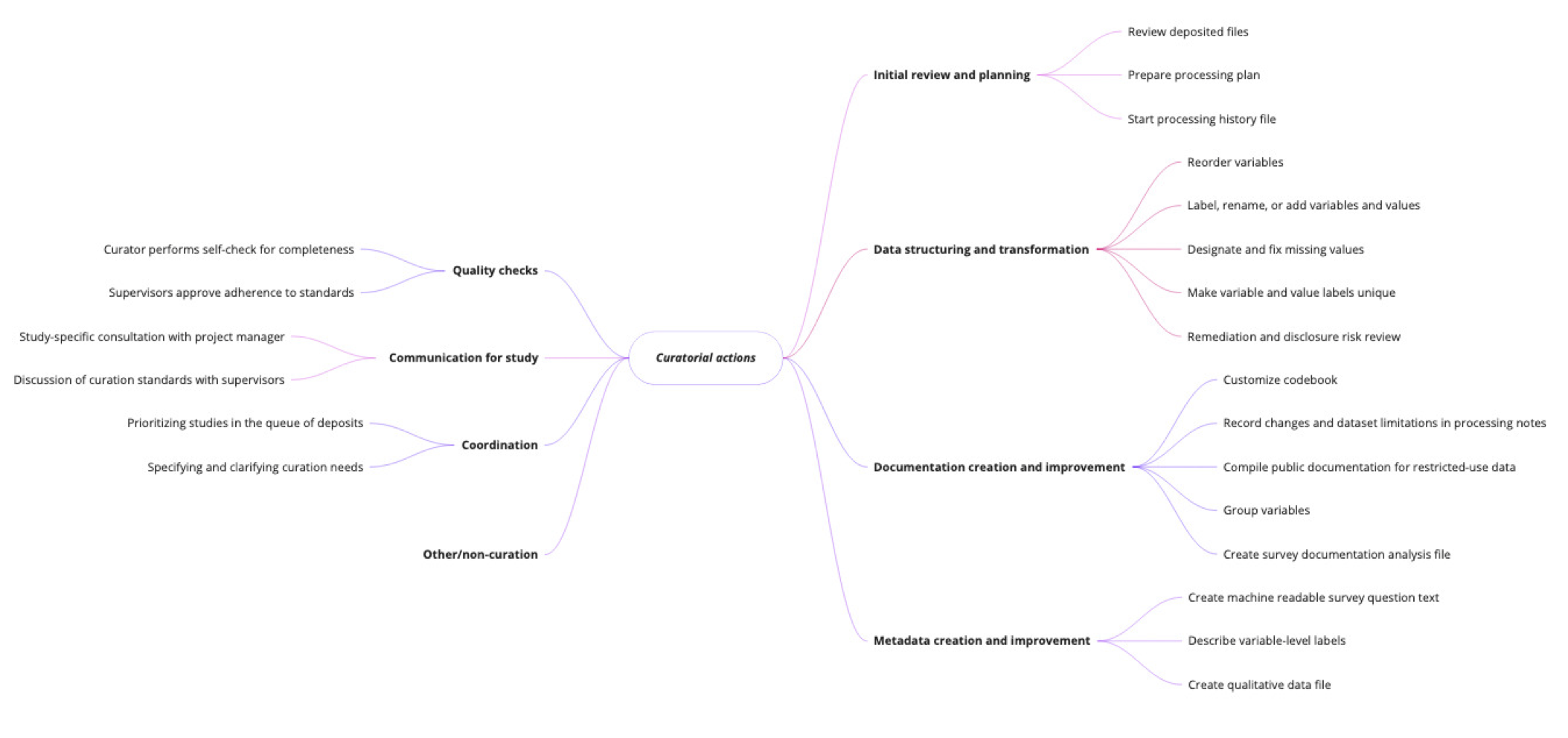}
    \caption{High-level curatorial actions that occur throughout the curation process}
    \label{fig:taxonomy}
\end{figure}

In the interviews and Jira tickets, we found a consistent, overlapping vocabulary of actions describing typical curation work. We also found insights into the ordering and time spent on curation tasks (see Figure \ref{fig:taxonomy} for examples of each type of action). Table \ref{tab:curation_actions} summarizes the amount of time curators logged for each type of curatorial action. While there were some typical sequences in which actions are performed, curators describe considerable variability in their own day-to-day work, rooted in their specific preferences, expertise, and craft knowledge. 

In the subsections that follow, we first describe the core high-level curatorial actions that are undertaken at ICPSR. These expand prior accounts of the work of data curation, particularly in CSCW and data science, where it’s described as one small part of a process. Then, we describe how curators rely on their craft knowledge to navigate the “workflow” dictated by these actions. In doing so,\textit{ we show how best practices and craft practices are deeply intertwined.}

\begin{table}[h]
  \caption{Time spent on curation actions (Feb. 2017 - Dec. 2019)}
  \label{tab:curation_actions}
  \begin{tabular}{lll}
    \toprule
    Action & Total hours logged & Percentage \\
    \midrule
Communication & 3,249 & 7\% \\
Data transformation & 12,363 & 26\% \\
Documentation & 3,094 & 6\% \\
Initial review and planning & 5,778 & 12\%\\
Metadata for study & 2,669 & 6\% \\
Non-curation & 6,641 & 14\% \\
Other & 1,157 & 2\% \\
Quality checks & 13,075 & 27\%  \\
  \bottomrule
\end{tabular}
\end{table}

\subsection{High-level curatorial actions}
\label{hands-on-work-with-data}

\emph{Initial review and planning.} Data curators at ICPSR typically begin their
curation of a deposit by reviewing deposited files and metadata and
\emph{developing a processing plan} -- an outline of planned curation
tasks, depending on a dataset's designated curation level. More detail
about curation levels at ICPSR can be found in Section \ref{triangulating-with-jira-tickets}. These plans are developed by curators and reviewed
by curation supervisors, who answer questions, troubleshoot, and
generally advise along the way. In recent years, more initial review and planning actions have also been
recorded for higher levels of curation (Curation Levels 2 and 3), suggesting that relatively more attention may be dedicated to developing curation plans at higher levels
of curation. Initial review and planning accounted for 12\% of curation time over our study period.

Early curation work also includes \emph{disclosure risk
review (DRR),} in which curators evaluate the risk that publishing a
dataset might pose to research participants and identify appropriate
mitigation steps. Curators described DRR as a critical
way in which they add value to a dataset -- both because of the
anonymization it provides, and for the thorough oversight the DRR
represented. 



\emph{Data structuring and transformation.} This category of curatorial
work includes the most ``technical piece'' of curation {[}CU015{]}: the direct work with the
dataset itself to make it easier to use, share and archive. Data
transformation tasks include designating missing values (e.g., assigning
metadata to values such as ``no response'' and ``not asked''); adding
question text (inserting the survey questions verbatim); transforming curated
SPSS data files into other statistical packages (e.g., R, SAS, etc); and creating documentation (PDF codebooks and XML metadata files). Datasets sometimes
are split into multiple, more usable parts (for instance, into smaller
file sizes, or commonly used file formats), or merged into single files
from multiple sources. This data structuring entails more than just
mechanical reformatting; as one curator describes, ``a lot of times,
especially in the larger datasets, there's a lot of pieces to put
together that I think when we make those connections it makes it easier
for users to use the data.'' {[}CU023{]} Considerable expertise and
judgement are needed to structure and transform data well. Data transformation was comparatively time consuming, taking up 26\% of curation time over our study period.

\emph{Metadata creation and improvement.} Curatorial work includes the
creation of records that will be queried by users within ICPSR's online
repository. Metadata development is seen as a distinct task; where the focus of data structuring is on making the dataset usable in
and of itself, the focus on creating metadata is on supporting search
and retrieval of datasets. Curators saw metadata as particularly
important because it's ``the first line'' of access, {[}CU017{]} the first
thing users see. Metadata improvements include drafting or
revising a dataset's description; copying and refining metadata from the
initial data provider, such as data collection dates; creating question
text (i.e., writing out the full list of questions in the survey
instrument that generated the data); and defining variable-level labels
(i.e., creating a data dictionary that spells out what each data
variable represents). Metadata work accounted for 6\% of curation time. This work is both qualitative and technical;
curators must have skill manipulating metadata standards to create these
records, and they need to have the experience to understand what context
is necessary and helpful for data reusers to include in metadata records.

\emph{Documentation creation and improvement.} In addition to creating
metadata records, curators also develop other forms of documentation
about the datasets. This includes creating processing history files,
codebooks, which include information for each variable in a dataset,
documentation of major changes made to the data, and compilation of any
additional documentation archived by the data producer. Documentation
accounted for 6\% of curation time logged. More instances of
documentation were recorded for curation activities in topical archives
than in the general archive.

\emph{Quality checks (QC).} These include
checking data and metadata files for completeness, confirming that the
work done to a dataset aligned with the Jira request, comparing the work
done to the processing plan, and confirming adherence to ICPSR's
guidelines and protocols for curation. The vast majority of studies
include quality checks, which was the major category of curation action we
detected in our analysis of Jira ticket worklogs, accounting for over 27\% of
curation time logged. These quality checks are performed by a second
curator to provide an extra level of review. 

Beyond designated quality checks, stakeholders discussed the value that
curation provided in ensuring that the data was of high quality overall.
Project-related communication is one mechanism for ensuring high quality
curation, which accounted for about 8\% of curation time logged in Jira tickets. This
includes catching issues and addressing complicated challenges with data
that data producers did not, providing high quality documentation about
data and the data curation process, setting and meeting goals for data
release that match depositor expectations and deadlines, and being a
source of consistent, vetted data. A curator described the last item in this
way:

\begin{quote}
Our work, I think, is pretty impactful and benefits the community
because the work we put in {[}will{]} rule out all the troubleshooting.
We look at the data, compare it to the documentation, and then do these
things to make sure everything's consistent. If there's any problems, we
either resolve with the PI or we have our own solution for it, so once
you get your hands on the data, there's nothing really in question for
the most part. {[}CU018{]}
\end{quote}

\subsection{Using craft to work the curatorial workflow}\label{curators-and-the-craft-of-data-curation}

The previous section outlined the tasks involved in the technical work of data curation. Actually accomplishing that work, however involves craft. Specifically, curators organize their work by first developing a gestalt, abstract mental representation of the data to envision what the final released dataset will entail; they then use their judgement and expertise to interpret standards, creatively come up with solutions, and thereby achieve a standard outcome in unstandardized ways. Paraphrasing one curatorial supervisor, ``I'm the curator: I do anything needed to make the data archivable'' [CU015].  We describe these two aspects of curatorial craft practices in the following subsections.

\subsubsection{From abstract representations to fit-for-use}


Curators approach a new dataset by getting the gestalt of the dataset: understanding the whole of the dataset as beyond the sum of its parts, as well as how these fit together in order to assess the feasibility of the processing plan and to envision the archivable and disseminated dataset. Several curators described getting the gestalt:
\begin{quote}
“I don't ... I do find the plan useful and it has to be done and there are things that it walks you through that can help you find missing things or problematic things. But I also rely heavily on just running the frequency output on the data and just scrolling through it. ... But I have found,  many times I have found things there that I wouldn't have seen otherwise. So I find that very important. And then I just ... yeah, between the plan and this check on my own, I find out what I need to do with the data and the documentation ... I'm not always very linear in how I work with this. I tend to do a lot of poking around...” (CU017)
\end{quote}
\begin{quote}
``So myself personally, I'll try to work with the data first, make some data manipulations or changes after I've gone through and read the documentation that's been provided, and get a good sense of what's going on with the study. But I won’t... So I'll read through everything and I won't fill out the metadata at that point because I still like to be able to go through and actually work with the data before I fill out the metadata that explains the collection in some more detail.'' [CU023] 
\end{quote}
Much of the process of getting a sense of the data is done with the user in mind. Curators think about how the dataset would need to be structured or documented in order to be fit-for-use for a range of users. Multiple curators describe customizing their work, or making decisions with the goal of supporting users’ access, essentially envisioning themselves as the ``first user'' of a dataset, and ``trying to figure out everything that a potential user would want to know and to make sure the archive version that we release is as complete as possible'' [CS009].
The curators’ goals are to answer any questions future users might have about the data: "We obviously can't anticipate everything, but we try to say, okay, if I was just picking this up, what would I need to know about it that maybe I don't have in a quick glance?" [CS010]. They also want to let users know about known issues "so they don't have to dig through it themselves to figure it out" [CS010]. 

Curators anticipated other types of users and user questions. For example, one  described tailoring datasets to a range of users: "Our data needs to be easy enough to use for the most novice user, but sophisticated for the more advanced user as well. And I think that that can happen by doing the details, making it easier" [CU017]. A curatorial supervisor considered how the dataset could be represented through crafting good metadata in supporting of use: 
\begin{quote}"And then even in our metadata. So I recently was talking to someone because I could tell them that the metadata ... Like the way they have worded was too internal-facing. I said, "That's not going to mean anything to users." If they don't understand, they're not going to hear about it. So we need to make it in a way that it's going to be something that makes sense to them. And is useful for them." [CS014]\end{quote}
And a third curator specifically linked the craft and subjectivity of curation with being able to conceive of how different users might perceive a dataset. \begin{quote}"I look at curation as, you know those technical aspects, there’s do’s and there’s don’t’s, right and wrong. There's also some, I like to say artistry, subjectiveness do it and how I might perceive a group of people wanting to view the data. Another person might see it differently." [CU016] \end{quote}


The gestalt techniques used by curators manifested itself differently in different cases. CU017 (above) expressed creativity in the information they chose to highlight for users. Several curators described differences in the order in which they approached curation tasks for a data study, or the time or level of detail they devoted to certain activities over others such as developing the processing plan. This ability to assess the present data, conceive of the path to a future state and conceive of a future representation is one aspect of craft exhibited by the curators. 

\subsubsection{Achieving standardization through judgement} Standards play a large role in data curation: at ICPSR, these include internal ``house'' standards set by ICPSR itself, and external standards developed by the broader community, such as metadata standards or preservation best practices. However, the application of standards is far from rote. Curators use their expertise and make judgements about when, how, and why to apply standards throughout the curation process.
Several participants said that because the data that ICPSR receives is just too diverse for strict standards to be feasible, and curators must rely on their craft knowledge and expertise to navigate "gray areas"  [CS012]. This can happen with "unusual data sets" in unique formats or with idiosyncratic structures [PM028], or for particular archives with distinct user communities, or for instances where a PI has requested what one participant called ``a la carte'' curation, where they do everything from one level, plus one task from another [CS009]. One supervisor said there are multiple workflows that stem from agreements with PIs, and therefore, one singular workflow isn't possible:
\begin{quote}
"I do think that we will always have more than one workflow just because sometimes the way proposals have to be written ... Sometimes project officers have a certain thing in mind. So, I think I mentioned earlier some PIs want a lot of involvement in disclosure review or the changes that we're making. So, for the demography archive, for pretty much most, if not all studies, we basically list out the examples of things that we're changing, and send it to the PI for approval. So, that's a different work flow than normally we just kind of can proceed. And then there's an archive within the criminal justice archive where the analysts want to use the data to do their analysis and then publish reports before we release the data. So, we have a process there were we do the first quality check then send it to them for review. They might send changes back to us, they might spend the next six months analyzing data. And so, we may not return to that study to make changes and/or do the second quality check and release it until months, or even a year later." [CS001]
\end{quote}

Some curators expressed frustation with standards. Long time curators particularly viewed the standards as living, malleable tools that change over time, rather than as unbreakable rules. Some went so far as to say that standards could be an obstacle to their work, because they interfered with their preferred way or working. One curator felt that ``... some of the standards that we have get in the way of it when they're supposed to help it.'' [023] A long-time curator observed: \begin{quote}They're getting better. Previously, there was just a lot of unanswered questions in the document. A lot of ambiguity. … Sometimes I'm a little outspoken just because I've been around and... We call them standards.” [CU017]\end{quote}This curator went on to say:
\begin{quote} "I don't necessarily agree that they’re standards. They're somebody's opinion that got put down and then set as a standard. It's somebody's preference then they made it a standard, especially on sentence case for variable labels that one's like... Those aren't standards. That's somebody's preference and I don't like it, because I would just do it differently. Not that my way is right and they're wrong. It's just different ways of thinking. If I was the one creating that standard it'd be different because my preference and my thought process is different." [CU017] \end{quote}

Several curators discussed applying standards creatively and flexibly in service to users, extending the curator’s focus on the user discussed previously. One curation supervisor acknowledged the importance of creativity to sidestep standards when the user was not served: 
\begin{quote} "So yeah, keeping those standards in mind, it's just sometimes you have to be creative. If there's something that you know users need to know who, put it in the summary field, don't put it in the collection notes, make it more visible. Maybe your tools don't allow us to make it as visible as we'd like. But there's always a way." [007] \end{quote}

Curators also were aware that the applying the standards involved judgement calls and could be self-reflective on their comfort level in making some types of these calls:

\begin{quote}”... so there's a lot of judgement calls involved despite all of our efforts to write up standards and follow them, again it's human produced data and documentation. And there're still judgement calls to be made from grammar and spelling and capitalization to more serious matters. So there's a number of judgement calls that I feel comfortable making. such as those involving labels. But then on the other hand, if it's a really sort of hairy disclosure risk scenario, I would definitely check in with my supervisor on those.” [MICA 24]\end{quote}





In deciding to apply or not apply standards, curators use expert judgement, creativity, and skill to achieve a standard outcome. One curation supervisor described their work as helping maintain the standards among curators but not setting or enforcing them. [012] Curators were also self-reflective about the standards and considered the data themselves as well as potential users when arriving at solutions that fell outside a ‘normal’ application of the standards. Whether the curators were more respectful or skeptical of the standards, many discussed instances where the standards fell short of achieving a dataset fit-for-use.





\subsubsection{Organizing curatorial actions}
Though there is a \textit{common} sequence of curatorial actions at ICPSR, there is not a strict workflow; processing plans outline the work that's needed at a high level, but not how it should be carried out. Curators use craft knowledge to sequence their technical work, and to customize their work practices to the dataset at hand or their own preferences. As one supervisor described,
\begin{quote}
 "We always say that curation isn't a linear process. There are a set of tasks that it makes sense to do in a particular order sometimes... I mean, we try to leave it up to the curator to what it works best for them because everyone has different ways of curating so... some people like to do metadata first, some people do that last, some people want to make all these data edits right away, some people want to focus on peripheral stuff. We try to get the processing plan done as soon as possible just because that helps expose all of the other issues that we might need to go to the project manager or the PI about. And that gives us more information about if we need to prioritize something particular." [CS013]
\end{quote}

The curators themselves described considerable variability in how they ordered their worked:

\begin{quote}
"I'm very collective, and it's not always the case, but just as I would prefer to be working on multiple curation projects at a time and instead of just focusing on one all day, every day until it's done, I like to have two or three going, if possible, just to break it up, break up my day, break up my focus. I also have that same approach for the tasks. So I might jump between different things." [CU016]
\end{quote}
\begin{quote}
"Well the prioritization is to complete the plan and the disclosure risk worksheet. So that is where I start. And in completing those, I set the agenda, so to speak, for where it goes next." [CU024]
\end{quote}
\begin{quote}
“I tend to start with the data, I tend to leave metadata to the end because I often find ... it could go either way, right? You could do the metadata to inform how you approach the data, but I often find that going through the data, going through the questionnaire lets me fill in the metadata better. Yeah. So my first step is the plan and the worksheet, because that is the first part of the process, and there's checks involved with other people. And from there I usually tackle the data first. [CU024]
\end{quote}
\begin{quote}
"I definitely jump around." [CU015]
\end{quote}
The common thread throughout these different approaches is that curators draw on their own expertise to structure their work and days: they know what works best for them and how best to hone their attention for detailed, technical, and sometimes tedious work.

\subsection{Coordination in service of curation}
Above, we described how curators use craft practices to gain a gestalt understanding of a dataset's structure, and then to organize their own work. This work is not done in a vacuum, however; curators must also coordinate with other stakeholders at ICPSR to proceed with this work, and to clarify priorities. Because ICPSR's workflows resist standardization, curators and curation supervisors consult archive directors, project managers, data producers, and each other to ensure there is a consensus (if not agreement) on the best way to approach curating a study. This occurs throughout the curation process.  For example, coordination occurs early on to determine where a study is placed in the curation queue and identify the level of curation. It can also occur later in the curation process if issues emerge requiring a decision about additional curation activities which are required to make the study fit for use. Acts of communication are captured in the Jira ticket worklogs, but the content is often vague. Our interviews elucidated the frequency and critical place of coordination in the curation process. These include the following: prioritizing studies in the queue; specifying how data will be curated; and monitoring progress and alerts.

\subsubsection{Prioritizing studies in the queue of deposits.} Curation supervisors manage a large queue of studies waiting to be processed and assess how, when, and to whom to assign them based on the priorities and funding available to the various topical archives at ICPSR. This assessment includes tight coordination with archive directors, data producers, project sponsors, and project managers. Curation supervisors factor in a project's budget, promised deliverables, relevant external deadlines, and the potential impact of the study's release to determine placement in the queue. Two archive directors described this balance of considerations:

\begin{quote}
...our funder really decides what to archive. I work with our project
manager to {[}...{]} ensure that the curation team is prioritizing our
data the way we want it prioritized. {[}...{]} And the project manager
ensures that those {[}...{]} priorities get communicated to the curation
team. {[}AD004{]}
\end{quote}
\begin{quote}
...I would say that feedback from the funders influences both the
curation levels and the priorities that we give to studies. So we do
coordinate with our program officer. {[}...{]} If there are certain
studies that are a high priority and that is something that we would
then incorporate into Jira and into the curation requests so that we can
adjust priorities and make sure that the highest priority work gets
prioritized accordingly {[}AD029{]}.
\end{quote}
Project managers also communicate priorities to the curation unit. They
do this as a matter of routine through multiple, reinforcing channels:
project managers enter deadlines and rank relative priority in Jira
tickets (e.g. Highest, High, Medium, Low), and they hold quarterly
meetings with curation supervisors to ``talk about the queue for a
particular quarter'' (PM025). However, as several curation and project
management staff noted, priorities change, and the communication often
involves significant back and forth:
\begin{quote}
There's a lot of back and forth in terms of what their priorities are
versus what we feel we can reasonably accomplish in a given timeframe.
And so sometimes that can get a little tricky. So in terms of like, if
they say we have this and this, we're going to ask them which one is
more important to them? I'm not going to try to figure that out, if I
can assign them both I will, if I can't I'll make sure it's their
highest priority. {[}CS010{]}
\end{quote}


\subsubsection{Negotiating levels of curation} 
Though the project managers initially define the work expected by choosing a curation level (1, 2, or 3), curators sometimes find that a given dataset needs more or different curation than originally planned. When this happens, they (and curation supervisors) must negotiate up and down the organizational chart to come to an agreement about how curation will proceed.
Curators and curation supervisors supervisors negotiate with project managers, who coordinate with archive directors, who sometimes coordinate with PIs. A curation supervisor described the negotiation process that can be involved in curation level clarification:

\begin{quote}
{[}...{]} We have curation levels and the project manager reviews those
levels and says, "Okay, I want this level of curation." And then we
would review it to see if that's accurate. So that would be like a
collaboration between the supervisor and the curator. So when {[}the
curators{]} do their processing plan, they may identify, "Hey, they're
asking me to do something that isn't in this level." Or ``they're asking
me to do things but it should be like a level up or down.'' And then we
also do a review of the plan and then we assess as well. {[}CS014{]}
\end{quote}

We note here that the project managers may not necessarily get the same gestalt view of the data as the curators, they trust the curators' view in further structuring work. As curation proceeds, and curators get into the data, they often discover things that suggest several possible courses of action that can
prompt discussion with curation supervisors, project managers, archive
directors, and the data providers themselves. One curator described
working with their supervisor to make final decisions. Though the curator ultimately defers to the curation supervisor, there's still a conversation about potential options:

\begin{quote}
If we've identified an issue or something, {[}...{]} I might give
{[}the supervisor{]} some options and then we talk about it a minute,
and ultimately {[}...{]} I let her decide as the supervisor, especially
when it comes to things on how to address confidentiality things. Those
are definitely things that supervisors would like to have their approval
on before things get out. It lessens the responsibility in a way on us
by having that supervisor, or someone who's in charge, being able to
make the final decisions {[}...{]} It's good to have, I think, other
people's opinions on those things. Yeah, for my supervisor, it's
definitely a conversation of, "Here's some possibilities of what we
could do." {[}CU015{]}
\end{quote}
Thus, there is some tension between respect for the curators' expertise and deep knowledge of their data, ICPSR's standards and decision-making hierarchy, and the overall budget for a project. 


\subsubsection{Monitoring progress via alerts, and navigating varying degrees of visibility} One mildly controversial method of facilitating coordination is through the use of Jira tickets to monitor and record progress on a project. Curators, supervisors, and project managers communicate about the study via Jira ticket comments. Curators receive an alert every time tickets are updated, and the tickets act as a running log of the work performed on the study. Though project managers found Jira to be generally helpful, some curators characterized Jira as annoying or overwhelming. The deluge of alerts and documentation also made some curators feel micromanaged, and as if Jira was keeping a running log of their work for their supervisors to review at any moment. As with any representation of work, however, the Jira tickets can be more or less precise. As one curator noted:
\begin{quote}"So I have trouble ... occasionally I have trouble keeping up with the ticket, the JIRA ticket, where we're meant to tick off things as we go because I'm kind of doing a little bit of everything at once because every part has information you need that affects other parts. I often find myself quite close to the end and I'm like, "Oh shoot, I have to go update the ticket." [CU024]\end{quote}

The curators' feeling of sometimes being overly visible is mirrored by other concerns about curatorial work being under recognized, or that curatorial work was insufficiently visible. For instance, one project manager said that they felt curation skills were invisible to those who are more removed from it: \begin{quote} I think that generally a lot of project managers and directors think that it's simple, simple syntax being applied but some of the challenges that come up while curating data can be quite complicated. It can take a certain level of skill. {[}PM026{]}
\end{quote}

The centralization of the curation function has taken the curators out of the individual archives, and the implementation of a single, shared standard has altered their practice to align with the organization, rather than with a single archive within ICPSR. In this new arrangement, ICPSR staff interact with curators at discrete points in the curation process, but few interact throughout the process. Therefore, there is less opportunity to see how curators use complex representations to envision the data as fit-for use and how they use judgement and creativity to achieve standardized outcomes for data. More coordination via intermediaries -- whether Jira or project managers -- becomes necessary to support curatorial work.

\section{Discussion}

\label{discussion}

\subsection{Understanding data work: more than just technical}
\label{defining-curatorial-work-more-than-just-technical}
One of the motivations of this study was to create a finer-grained understanding of data work, specifically focusing on curatorial actions. We developed a rich description of data curation work at ICPSR -- one that goes beyond the technical, procedural work with data and metadata to include the expertise-driven decision-making involved in crafting data, and the coordination required to develop a consensus around curatorial priorities and activities. Our participants have shown that data curation is neither ``magic'' nor ``janitorial'' work \cite{DIgnazio2020-es,Rawson2019-hq,Owens2018-gm}, but rather, is the result of technical skill enacted through craft practices. Indeed, we find that staff members in all roles bristle against characterizations of curation as something rote or mechanical. Curators do what what needs to be done to achieve the outcomes of a standard -- even when not necessarily \textit{following} a standardized workflow. This requires significant collaboration with other stakeholders in the data science workflow. Thus, the workflow is achieved but much of the actual work that made that happen disappears.

Our work makes two main contributions to understanding data curation, and thereby data work. First, the description of ``hands-on'' technical tasks we provide in Section \ref{hands-on-work-with-data} expands an existing body of literature describing data curation practices in different contexts. Understanding different data (curation) practices is critical for building infrastructure, software tools, and ontologies that capture disciplinary contexts, and for educating curators. For instance, \citet{Chao2015-sk} developed the Data Practices and Curation Vocabulary, which describes how a community (in that case, earth scientists) defines data curation. Comparison of our two frameworks reveals that ICPSR has much more detailed quality check protocols, and ICPSR's curators spend considerable time on tasks like ``adding question text'' that simply are not needed in the earth science fields. The diversity of curatorial actions shown in just these two papers highlights the need for further research into the specific curatorial workflows and communication regimes in different scholarly settings. It is well understood from research on data practices that there are significant domain differences in curation needs \cite{Faniel2019-mt,Akers2013-ea,Witt2009-nk,Cragin2010-lw}. Yet models of data curation rarely account for this diversity of practice, or provide guidance in how to navigate them.

Second, our work shows the vital role that craft practices play in successfully organizing curatorial work and applying and navigating standards. 
\textit{In data work, we see craft manifesting as the ability to develop an abstract, gestalt representation of a data product and then envision how to make changes to that data product so that it is more fit-for-use. This work involves following best practices and creating a standardized product, but not necessarily following a standardized workflow.} Furthermore, the kind of data curation carried out at ICPSR requires significant collaboration and consultation with other stakeholders. This extends prior work on craft in technical settings in the
CSCW literature, most recently discussed by
\citet{Muller2019-gi} in their summary of craft in the context of
data science, as well as a more recent focus on craft in the LIS literature by \citet{Owens2018-gm}. Muller and co-authors
summarized key themes regarding craft in CSCW, including ``Conversation
with materials: Through the conversation with materials, there is often
a sense of intimacy with materials and media'' and ``Control:
Craft-workers labor at an intersection of control and
unpredictability.''

At ICPSR, we see clear alignments with some of Muller et al.'s account
of craft. Curators repeatedly emphasized the importance of the
``conversation with materials'' in their work through repeated
descriptions of the contingency of their workflows and specific tasks. 
Likewise, ICPSR curators exist at the intersection of ``control and
unpredictability'' -- they are constrained into somewhat narrow roles by
ICPSR's organizational structure, yet must navigate unpredictable and
unique curation challenges for each dataset with stakeholders
throughout and outside of ICPSR. Our research further shows how craft practices ``fit'' into best practices and other standards for working with data; \textit{in short, we find that craft practices are necessary to enact best practices.} It's well understood that data standards can vary in their application and results based on variations in how they are enacted by a group \citet{millerand_trajectories_2009}. Yet at ICPSR we see \textit{a standardized result arising from the nonstandardized application of standards via craft practices}. By giving curators the freedom to rely on their own skill to structure their work and make decisions, ICPSR is able to truly rely on them as the human-in-the-loop.

Accounting for the role of craft and expertise in data work is important in designing effective data workflows, training data workers, and in better supporting data workers in showing the impacts of their work. We expand on this further in section 5.3 We argue that this view raises important questions for the practice of science (data science, social science, etc.), such as:  How do notions of ``craft'' complicate the development of data curation pipelines to support complex data science applications or support repository infrastructures that automate curation? We begin to consider the latter question in the following section. How does understanding the craft invovled in data work support data workers in gaining credit for their work and it's impact? We address this question in section 5.3


\subsection{Coordinating work in data curation: complicating ``workflow'' or ``pipeline'' views of data science and curation}
\label{coordinating-work-in-data-curation-complicating-workflow-or-pipeline-views-of-data-science-and-curation}
One of our primary findings is that data curators must structure their own work within the context of their organization's structure and job descriptions and constraints. In this way, they and other stakeholders ``work the workflow'' and navigate across standards and up and down the organizational chart; they gain a gestalt view of not just the data at hand but also of the organization as a whole. Coordination and communication are key in this. In identifying coordination and craft practices as important parts of data
curation work, we complicate not just solely technical accounts of data
curation, but also ``workflow'' or ``pipeline'' conceptions of data
work. By ``workflow'' views, we mean conceptualizations of data curation
as a sequential process, easily represented by a UML diagram or similar
technique. These representations are quite common in CSCW and the
information sciences, where they are used to model curation processes at
a higher level
\cite{Zhang2020-bf,Muller2019-gi,Kross2021-gz,Johnston2014-tc},
or the plethora of data/digital lifecycle models in the digital curation
literature
\cite{Higgins2008-og,Faundeen2013-ww,noauthor_undated-tj}),
or to capture detailed change logs and provenance chain of a dataset
\cite{Thomer2018-na,Goble2008-kc,Zhao2012-ja,Goble2010-nw}.
The models are common because of their utility; they represent complex
processes in a way that is digestible, and they can act as boundary objects
that help communication between disparate groups of stakeholders
\cite{Dourish2001-hs}.

However, our work here underscores that data curation is more than the sum of its parts, involving much
more than the objects that are curated; it is also a process in which
distributed knowledge management decisions are made to facilitate
information reuse \cite{Ackerman1999-fz}. Our research
supports the notion that data curation is a highly collaborative process
occurring across a distributed system over time. While some curation
actions tend to occur in sequence, important components of curation
work, like quality checks, are performed in parallel or iteratively
throughout the curation process. Project-related communication is also
embedded in all other curatorial actions, making it difficult to
delineate. A closer look at project-related communication reveals the
importance of discussion and delegation in curatorial work; for example,
supervisors and curators often discuss how best to mitigate disclosive
variables on a case-by-case basis, following risk minimization
heuristics rather than hard rules. And though coordination strategies
such as \emph{Prioritizing studies in the queue of deposits} and
\emph{Specifying and clarifying how the data will be curated} may seem
like they could fit neatly into a workflow diagram, in reality, they
require a meta-level understanding of the curation workflow itself to
proceed. Articulation work is needed to navigate a data science workflow
\cite{Neang2021-jv,Thomer2018-oq}, yet this labor can,
somewhat ironically, be obscured in workflow-centric views. We want to
be clear: we are not trying to discourage or dismiss workflow-based
explorations of data work. Rather, we want to note the importance of
continued, rich exploration of what goes on in and around each ``box''
of the diagram -\/- lest we obscure that which we wish to reveal.

\subsection{Revisiting the invisible nature of data curation}
\label{revisiting-invisible-curatorial-work}

In our prior work, we have argued that hiding curation makes it harder to plan, prioritize and fund curatorial activities (anonymized for review).  Additionally, by rendering their work invisible to outsiders, curation can hide curators' value and impact. Our interviews verified this latter point, in that curators -- and even their managers, to a degree -- described some concern that their work was not truly seen or appreciated. Invisibility can make it harder for these data workers to advance in their careers, lobby for salary increases, and participate fully in their fields. Our work here reveals some tensions, though, in making curatorial work totally visible. Below we discuss both the visibility and hypervisibility of curatorial work at ICSPR.

The craft and coordination in curatorial work at ICPSR are mostly invisible to data users. The public datasets hide the work that went into their creation precisely because they are standardized \cite{Plantin2019-fs}. Even the documents that emerge from curation hide aspects of this work; while the Jira tickets contain descriptions of the high-level tasks, they do not provide a full account of the curators' labor and decision making process. The existence of data curation standards makes the work seem routine even though all our interviewees recognize, to varying degrees, that curation requires technical skill, flexibility, and coordination. 

At the same time, some aspects of curators' work is hypervisible within ICPSR through Jira tickets and other documentation. The Jira ticket worklogs and comments,
especially, serve first to coordinate work and then to
document it.  And while Jira can document their labor and
decisions, making their work visible, it can also open curators to
negative side-effects such as micromanagement. For instance, Jira tickets make it
possible for more powerful colleagues (e.g., archive directors) to
monitor curators' work. As \citet{Suchman1995-zr}
and \citet{Yates1989-fo} pointed out years ago,
technologies that help workers coordinate locally can become mechanisms
of global control by enabling surveillance and proscription. 
Thus, not all invisible work should be made visible. 
\citet{Bishop1999-tv} uses Weber's concepts of ``status
contract'' and ``instrumental contract'' to understand the changing
relationships between employers and employees. She notes that status
contracts -- those that are about our relations to one another rather than
our performance -- often rely on the trust that results from these
relationships, and not from formal articulations of the work. At ICPSR, we see evidence of this status contract; by and large, those higher in the organization respect the skill and expertise of their curators. There is an understanding that some aspect of data work will always be invisible. The use of Jira tickets to monitor, however, threatens to replace this status contract with an instrumental one, in which the worker is valued for visible products. 

How does viewing curation as a craft impact this (in)visibility? When supervisors, project managers, and archive directors view and treat curation as a craft, it supports the status contract between curators and higher management. It appreciates this data work as skilled labor, and thereby "affords identity, status and a sense of connection to others in the enterprise and to the enterprise itself'' (\citet{Nardi1999-tu} citing \citet{Bishop1999-tv}). When we as data practices researchers, data science educators, and CSCW theorists argue for curation as craft, we, too contribute to the support of this status contract. Thus, recognizing curation as craft is important to supporting labor arrangements that do not render the worker invisible even when the work is.
A well known impact of the invisibility of curation work is that outsiders underestimate its costs and value, and, by implication, the value of curators. Work like curation that is conducted in the background is
often taken for granted. Recent efforts to surface curatorial contributions to scholarship via structured metadata \cite{Thessen2019-ps} or improvement of legacy data records \cite{noauthor_undated-gx} echo prior efforts
such as the Nursing Interventions Classification to make work visible in
efforts to legitimate both the work and workers
\cite{Star1999-fx,Bowker1996-ue}. Here, we are pushing to recognize the labor needed to organize, understand, and negotiate the tidy boxes on workflow diagrams -- and to recognize it's seeming ineffibility as important to preserve and respect.

\subsection{Implications for practice}
\label{implications-for-practice}
Better articulating the work and craft of data curation has several implications
for practice. First and foremost, understanding the complex role that different forms of visibility play in data work may help us design
technologies for users that move beyond reporting and surveillance
\cite{Suchman1995-zr}. As we described, many ICPSR
curators bristled at the constant use of Jira because it made them feel
hypervisible, monitored, and mildly harangued. We consequently ask:
given a view of curation as craft rather than rotely mechanical labor,
what changes might we imagine for ticketing systems like Jira -\/- ones
that might lead the management system to serve the curators as well as
their supervisors and managers?

Our work also has several implications for data curation training and
education. Within the information sciences, considerable effort has been put into designing data curation curriculum for budding information professionals; much of this has been highly focused on articulating different versions of data curation workflows, and describing data practices in different fields. The range of high-level curatorial actions we identified in section 4.1 contributes to this tradition. 

However,
the greater contribution of our work is the importance of training data
curators as craftspeople and not just technicians. As we quoted from \citet{Owens2018-gm} previously, a craftful
approach to curation is one that stays engaged with the ``unresolved''
and contingent aspects of curatorial work, and one that sees the
``inherent messiness'' of data work as a feature, rather than a bug. Our work helps more specifically identify the strategies ICSPR curators use to navigate this unresolved messiness, particularly in how they use gestalt approaches to see the dataset as more than the sum of its parts. Though more work would be needed to better understand this process, we believe it is a promising direction for further curriculum develop -- whether in data curation classes at the master's level, or online lessons in the vein of Data Carpentry (https://datacarpentry.org/).

One of the limitations of this study is ICPSR's unusual size and scope; they simply have a much larger and more well organized data curation team that many other peer institutions and archives. It is possible that lessons learned here will not translate well to smaller contexts or teams. However -- we believe this view of data work as grounded in craft practices could be important to explore elsewhere. How do craft practices differ in smaller organizations, or in teams where there are not dedicated curators?  For those that think of data curation as a "wrangling" or "munging" process, how could adopting a craft perspective help guide this work and make it more reproducible?

\section{Conclusion}
\label{conclusion}
Data curation is a critical component of data science, and an important aspect of data work. Obscuring the work of data
curation not only renders the labor and contributions of the data
curators invisible; it also makes it harder to tease out the impact
curators' work has on the later usability, reliability, and
reproducibility of data. In this paper we have made curatorial work visible through a case study of
data curation at ICPSR, a large social science data repository. We have
contributed a rich description of curatorial work at this site,
including a range of technical curatorial actions, and the craft and coordination needed to successfully enact those actions. We echo prior work calling for a
craftful view of work with data: curation requires not just a rote following of standards and protocols, but rather, a creative, on-going conversation with the data, with one's colleagues, and with one's community. Our work complicates "workflow" based views of data curation, in that we find ICPSR curators do considerable work that can't be easily visualized with a UML diagram, and indeed, rely on craft practices to work their worklfow.  We also find that ICPSR curators sit at an intersection between visibility and invisibility: their work is highly documented (and even monitored, to a degree), yet when they do their jobs well, it is invisible. Finding ways of selectively making curatorial work visible in service of curators will be key in supporting their work and professional development, as well as the development of data curation tools.

\bibliographystyle{ACM-Reference-Format}
\bibliography{CSCW-MICA-2021}

\appendix

\section{Interview protocol}
\label{appendix-a-interview-protocol}

\underline{Background}

\begin{itemize}
\item
  \begin{quote}
  How long have you been a {[}ROLE{]}?
  \end{quote}

  \begin{itemize}
  \item
    \begin{quote}
    What's your academic background?
    \end{quote}
  \item
    \begin{quote}
    Do you have any prior experience as a data manager, curator, etc.
    \end{quote}
  \end{itemize}
\item
  \begin{quote}
  I would like to start by hearing more about what your {[}ROLE{]} work
  at ICPSR.
  \end{quote}
\item
  \begin{quote}
  What is your role in the curation process?
  \end{quote}

  \begin{itemize}
  \item
    \begin{quote}
    Would you describe the chain of command? (e.g., Archive directors,
    curators)
    \end{quote}
  \item
    \begin{quote}
    How do you work together to make decisions?
    \end{quote}

    \begin{itemize}
    \item
      \begin{quote}
      How do you interact with project managers?
      \end{quote}
    \item
      \begin{quote}
      Do you feel involved in making judgement calls?
      \end{quote}
    \item
      \begin{quote}
      When a study seems like it falls between two levels of curation,
      who determines which level to assign it?
      \end{quote}

      \begin{itemize}
      \item
        \begin{quote}
        What factors are important to this determination?
        \end{quote}
      \end{itemize}
    \end{itemize}
  \item
    \begin{quote}
    Can you describe your overall workflow to me?
    \end{quote}
  \item
    \begin{quote}
    Is your work specialized to an archive/domain/data type?
    \end{quote}
  \end{itemize}
\item
  \begin{quote}
  How much curation do the datasets you work with typically need?
  \end{quote}

  \begin{itemize}
  \item
    \begin{quote}
    Do they tend to arrive in the same state?
    \end{quote}
  \end{itemize}
\item
  \begin{quote}
  Would you tell us a bit about the scholarly community that uses your
  archive?
  \end{quote}
\item
  \begin{quote}
  What type of relationship does the archive have with the scholarly
  community reusing its data?
  \end{quote}
\end{itemize}

\underline{Curation}

\begin{itemize}
\item
  \begin{quote}
  {[}If applicable{]} What types of interactions do you have with the
  curation unit?
  \end{quote}
\item
  \begin{quote}
  How involved are you in curation decisions?
  \end{quote}

  \begin{itemize}
  \item
    \begin{quote}
    When does this occur (grant proposal, initiation of a grant/project
    planning, before/during data sharing)?
    \end{quote}
  \item
    \begin{quote}
    {[}If applicable{]} Were you involved in recent decision to make
    ICPSR curation workflows more systematic? If so, can you tell us
    what led to that decision?
    \end{quote}
  \item
    \begin{quote}
    How have recent changes to curation workflows at ICPSR changed your
    involvement in curation decisions, if at all?
    \end{quote}
  \item
    \begin{quote}
    Is there a formal process?
    \end{quote}
  \item
    \begin{quote}
    Are these decisions always easy to implement?
    \end{quote}
  \end{itemize}
\item
  \begin{quote}
  Which curation activities add the most value to your archive/datasets?
  \end{quote}

  \begin{itemize}
  \item
    \begin{quote}
    Why do you say this?
    \end{quote}
  \end{itemize}
\item
  \begin{quote}
  How do you prioritize different curatorial activities?
  \end{quote}

  \begin{itemize}
  \item
    \begin{quote}
    How do you know when a dataset is ``done'' being curated?
    \end{quote}
  \item
    \begin{quote}
    How involved are you in making judgement calls (e.g. between levels,
    between different curatorial actions)?
    \end{quote}
  \end{itemize}
\item
  \begin{quote}
  Are the curation levels well defined?
  \end{quote}

  \begin{itemize}
  \item
    \begin{quote}
    Do you think they work for most studies?
    \end{quote}

    \begin{itemize}
    \item
      \begin{quote}
      Why or why not?
      \end{quote}
    \end{itemize}
  \end{itemize}
\item
  \begin{quote}
  How has your job changed since the curation reorganization? {[}Tailor
  to ROLE and BACKGROUND{]}
  \end{quote}
\item
  \begin{quote}
  Do your data reusers or designated community provide input into
  curatorial decisions?
  \end{quote}
\item
  \begin{quote}
  How has the curation provided by ICPSR changed the use or impact of
  your collections?
  \end{quote}
\item
  \begin{quote}
  Do you ever question the amount of curation planned for or being
  applied to a dataset?
  \end{quote}
\item
  \begin{quote}
  Is there additional/different curation you'd like to see applied to
  some of your datasets?
  \end{quote}
\end{itemize}

\begin{itemize}
\item
  \begin{quote}
  What metrics would you propose or like to guide the level of curation
  of data?
  \end{quote}
\end{itemize}

\underline{Impact and metrics}

\begin{itemize}
\item
  \begin{quote}
  What type of impact would you like your archive to have?
  \end{quote}

  \begin{itemize}
  \item
    \begin{quote}
    How close is the archive to achieving this goal?
    \end{quote}
  \item
    \begin{quote}
    Where would you like to see the impact of your archive in 5 years?
    \end{quote}
  \end{itemize}
\item
  \begin{quote}
  Are you aware of any metrics at ICPSR guiding the curation process?
  \end{quote}
\item
  \begin{quote}
  What metrics do you currently use to measure the impact of your
  collection, if any?
  \end{quote}
\item
  \begin{quote}
  What metrics would you propose to measure the impact of your
  collection?
  \end{quote}
\item
  \begin{quote}
  {[}If applicable{]} Do you plan or discuss your curation work with
  anyone else at ICPSR?
  \end{quote}

  \begin{itemize}
  \item
    \begin{quote}
    If yes, how do their comments impact your curatorial decisions?
    \end{quote}
  \end{itemize}
\item
  \begin{quote}
  {[}If applicable{]} Do you consider the potential impact of the
  dataset during curation?
  \end{quote}

  \begin{itemize}
  \item
    \begin{quote}
    If so, how?
    \end{quote}
  \end{itemize}
\item
  \begin{quote}
  {[}If applicable{]} How does your work add value to the datasets you
  curate?
  \end{quote}

  \begin{itemize}
  \item
    \begin{quote}
    Probe if not answering specifically: application of standards,
    metadata
    \end{quote}

    \begin{itemize}
    \item
      \begin{quote}
      Which ones?
      \end{quote}
    \end{itemize}
  \end{itemize}
\item
  \begin{quote}
  How do you see x (e.g., metadata, data cleaning, etc.) having impact
  on the datasets?
  \end{quote}

  \begin{itemize}
  \item
    \begin{quote}
    {[}If applicable{]} In what ways do the JIRA tickets document the
    curation work you've done?
    \end{quote}
  \item
    \begin{quote}
    {[}If applicable{]} In what ways do the JIRA tickets \emph{not}
    document the curation work you've done?
    \end{quote}
  \item
    \begin{quote}
    {[}If applicable{]} Do you find JIRA intrusive?
    \end{quote}
  \end{itemize}
\item
  \begin{quote}
  Is there a dataset that you've worked on that's had substantial
  impact?
  \end{quote}

  \begin{itemize}
  \item
    \begin{quote}
    If so, could you describe what it was and what impact it made?
    \end{quote}
  \item
    \begin{quote}
    What contribution did your curatorial work have on this dataset's
    impact?
    \end{quote}
  \end{itemize}
\item
  \begin{quote}
  How would your designated community define impact of the collections?
  \end{quote}
\item
  \begin{quote}
  What impact would the faculty that contribute data to your archive
  want the collection to have?
  \end{quote}

  \begin{itemize}
  \item
    \begin{quote}
    Do you believe that data sharing and reuse should be considered for
    promotion and tenure?
    \end{quote}
  \item
    \begin{quote}
    How broadly is this shared in the scholarly community served by the
    archive?
    \end{quote}
  \end{itemize}
\item
  \begin{quote}
  What kind of impact do you want your work to have? Your collections to
  have?
  \end{quote}
\end{itemize}

\underline{Reuse}

\begin{itemize}
\item
  \begin{quote}
  Do you consider data reusers during the curation process?
  \end{quote}

  \begin{itemize}
  \item
    \begin{quote}
    If so, what are the significant characteristics or properties (e.g.,
    information about the data that is important for effective
    preservation management or reuse) you think are important to capture
    to enable data reuse?
    \end{quote}
  \end{itemize}
\item
  \begin{quote}
  Do you interact with data reusers?
  \end{quote}

  \begin{itemize}
  \item
    \begin{quote}
    If yes, do their comments impact your curatorial decisions?
    \end{quote}
  \end{itemize}
\item
  \begin{quote}
  What are the greatest barriers in re-using collections from your
  archive?
  \end{quote}
\item
  \begin{quote}
  What kinds of input or questions do you get from data reusers?
  \end{quote}
\item
  \begin{quote}
  What do reusers tell you about the value of different datasets?
  \end{quote}
\item
  \begin{quote}
  What kind of reuse would you like to facilitate in the future?
  \end{quote}
\end{itemize}

\underline{Wrap up}

\begin{itemize}
\item
  \begin{quote}
  Do you have any questions for us, or about this project?
  \end{quote}
\end{itemize}

\end{document}